\documentclass[reprint,twocolumn,nofootinbib,showpacs]{revtex4}

\usepackage{amsmath}
\usepackage{amssymb}
\usepackage{epsfig}

\def\C{{\cal C}}

\def\be{\begin{equation}}
\def\en{\end{equation}}

\def\RR{\rm \hbox{I\kern-.2em\hbox{R}}}
\def\NN{\rm \hbox{I\kern-.2em\hbox{N}}}
\def\ZZ{\rm {{Z}\kern-.28em{Z}}}
\def\CC{\rm \hbox{C\kern -.5em {\raise .32ex \hbox{$\scriptscriptstyle
|$}}\kern

-.22em{\raise .6ex \hbox{$\scriptscriptstyle |$}}\kern .4em}}

\def\<{\langle}
\def\>{\rangle}

\newcommand{\EE}[1]{\mathbb{E}\left[#1\right]}

\newcommand{\Var}{\mathbb{V}\mathrm{ar}}
\newcommand{\Cov}[2]{\mathbb{C}\mathrm{ov}\left(#1,#2\right)}

\begin{document}

\title{Random cascade model in the limit of infinite integral scale as the exponential of a non-stationary $1/f$ noise.
Application to volatility fluctuations in stock markets}


\author{Jean-Fran\c{c}ois Muzy}
\email{muzy@univ-corse.fr}
\affiliation{SPE UMR 6134 CNRS, Universit\'e de Corse, Quartier Grossetti,
20250 Corte, France}
\affiliation{CMAP UMR 7641 CNRS, Ecole Polytechnique, 91128 Palaiseau, France}

\author{Rachel Ba\"{\i}le}
\email{baile@univ-corse.fr}
\affiliation{SPE UMR 6134 CNRS, Universit\'e de Corse, Vignola, 20200 Ajaccio, France}

\author{Emmanuel Bacry}
\email{emmanuel.bacry@polytechnique.fr}
\affiliation{CMAP UMR 7641 CNRS, Ecole Polytechnique, 91128 Palaiseau, France}

\date{\today}

\begin{abstract}
In this paper we propose a new model for volatility fluctuations
in financial time series. This model relies on a non-stationary
gaussian process that exhibits aging behavior. It turns out that
its properties, over any finite time interval,
are very close to continuous cascade models. These latter models
are indeed well known to reproduce faithfully the main stylized
facts of financial time series. However, it involve a large scale parameter
(the so-called ``integral scale'' where the cascade is initiated) that is hard
to interpret in finance. Moreover the empirical value of the integral
scale is in general deeply correlated to the overall length of the sample.
This feature is precisely predicted by our model that turns out, as
illustrated on various examples from daily stock index data, to quantitatively
reproduce the empirical observations.
\end{abstract}


\maketitle

\section{Introduction}
For several decades, random cascade models have been at the heart of a wide number
of studies in mathematics as well as in applied sciences. They were introduced to
account for the intermittency phenomenon in fully developed turbulence and are
involved every time one observes a multifractal (or a multiscaling) behavior
in the variations of statistical properties of some field across different scales.
Multifractal scaling is generally associated with the existence of a random cascade by which
small scale structures are constructed from the splitting of larger ones
and multiplication by a random factor. One clearly sees that such a scenario
necessarily implies the existence of a large {\em integral scale} $T$ where the cascade is initiated.
As emphasized below (see Appendix \ref{app:integralscale}), on a general ground, one can show that
the moment multiscaling behavior of the increments associated
with any multifractal field cannot hold over an infinite range of scales. It
necessarily involves a large scale $T$ above which scaling properties becomes trivial.
In turbulence this scale naturally corresponds to the injection scale, i.e., the time/space scale
where kinetic energy is injected into the flow \cite{BF}.
The main question addressed in this paper concerns the fields where a multifractal
behavior is observed without the existence of any obvious integral scale.
This is notably the case in empirical finance.

In quantitative finance, volatility is one of the most important risk measures since
it corresponds to the conditional variance associated with price
fluctuations at any time $t$ \cite{BBoPo}.
A well known stylized fact is that volatility fluctuations are organized into
persistent clusters. A huge amount of the econometrics literature
is devoted to the modeling of this volatility persistence. Among all the proposed
alternatives, the GARCH models \cite{Boll86}
and their extensions have been thoroughly studied.
The major drawback of such models is that, on one hand, their aggregation properties
are not easy to control and on the other hand, they cannot account for the long-range nature
of volatility correlations \cite{volfluct2,volfluct3}. This last feature translates
in the fact that GARCH parameters are often found to be at the borderline
of the stability region. This is the so-called IGARCH effect \cite{EngBoll86}.

Under the impetus of early studies of
Mandelbrot and his collaborators \cite{ManCalFish},
the notions of multifractals and random cascades have been proposed to account for the volatility
dynamics in many studies of financial time series (see e.g. \cite{turbfin,AMS98,MuDeBa00,Lux01,cfb}).
The class of continuous random cascades \cite{BaMu03} and in particular the MRW model,
provides a parsimonious class of random processes that reproduces very well most of stylized
facts characterizing the price return fluctuations \cite{MuDeBa00,BBoPo}.
Unlike GARCH models, these models are continuous time models (so they do not involve a discrete time
step) which aggregation properties are easy to handle since they possess some self-similarity properties.
Within this framework, various empirical estimations reported so far indicate
that the value of $T$ can vary from few months \cite{turbfin,AMS98}
to several years \cite{BacKozMuz11,mlmrw} (see Fig. \ref{fig9} below).
Even if it is well admitted that
a precise estimation of $T$ can be hardly achieved \cite{BacKozMuz11,mlmrw},
one can naturally wonder why one observes
such a large range in the estimated integral scale values.
Beyond the problem of the determination of $T$, a challenging question
remains to understand the {\em meaning} of the integral scale in finance.
Unlike turbulence, there
is no natural large scale that would obviously appear to be associated with
some ``source of volatility''.

The idea we propose in this paper is that such a scale does not exist (or
is formally ``infinite'') and that the volatility is
a non-stationary process. Let us notice that, within standard econometric framework,
many authors already proposed to explain the above mentionned IGARCH effect
by the non-stationary nature of volatility fluctuations:
these models include Fractionally integrated GARCH \cite{Bai96},
GARCH models with time varying parameters\cite{MikStar03,StarGran05},
stochastic volatility models with unit roots \cite{Hans95}.
Our approach is original in the sense that we
account for the non-stationary nature of volatility
fluctuations while remaining within the framework of multifractal processes.
Indeed, we will show that our model is such that every single trajectory,
for each finite time interval, can hardly be distinguished from the path of
a multifractal process where the {\em integral scale is precisely the length of the time interval under consideration}.
Our construction is written
as the exponential of a non-stationary $1/f$ noise
and is based on an extension of continuous random cascades based on infinitely scattered random
measure as introduced in Refs \cite{MuBa02,BaMu03}.
We show that this process is well defined in the sense of distributions
and cannot be distinguished from a continuous
cascade model (as the MRW process defined in \cite{MuDeBa00})
over any finite time interval far from the time origin.
We check and illustrate our results on some numerical simulations.
We then consider applications to stock index market data that are shown
to exhibit some ``aging'' behavior as precisely predicted by our model.

The paper is organized as follows: In section \ref{sec_mf} we make a brief overview
of multifractal models as they have been proposed to account for the volatility
fluctuations in financial time series. The construction of log-infinitely
divisible continuous random cascades as introduced in \cite{MuBa02,BaMu03} is also explained but
we mainly focus on the log-normal case.
In section \ref{sec_th} we show how one can some extend the former
cascade models in the formal limit $T \rightarrow +\infty$. The price to pay is
that the model is no longer stationary. However, this new model has appealing
properties since, in some sense, it reduces to a multifractal model over
any bounded time interval without involving any large scale parameter.
Our results are illustrated using numerical simulations.
In section \ref{sec_apps}, we address the problem of the model estimation
using a single realization. We then show that our approach is pertinent to account
for the observed volatility correlations from intraday to many year time scales.
In particular it allows one to understand the wide range of estimated integral scale values
reported in the literature.
We use the Dow-Jones daily data recorded over several decades
to provide evidence against the stationarity of the volatility process.
Concluding remarks and pathes for future research are provided in section \ref{sec_cp}.
Some technical results are reported in Appendices.

\section{Multifractal volatility models: a brief overview}
\label{sec_mf}

\subsection{Multiscaling}
As mentioned in the introductory section, multifractal models
have provided a family of stochastic processes that accounts very well for the main
statistical features of financial time series \cite{BacKozMuz06,BCF}.
In this section we recall the main results concerning random cascade models
and set the main notations. We refer the reader to Refs \cite{MuDeBa00,MuBa02,BarMan02,BaMu03}
for more mathematical details.

As first proposed by Mandelbrot et al. \cite{ManCalFish}, multifractal processes $X(t)$
with zero mean and stationary increments,
can be constructed through an auxiliary non-decreasing multifractal measure $M(t)$ as
\begin{equation}
\label{compound}
    X(t)  = B\left[M(t)\right]
\end{equation}
where $B(t)$ is a self-similar process (i.e. such that $B(\lambda t) =_{law}\lambda^H B(t)$)
in general chosen to be a standard Brownian motion ($H=1/2$). It results that the increments of $X$ and $M$ are related by:
\begin{eqnarray}
   X(t+\tau) \! - \! X(t) & \! \! \mathop{=} \limits_{law} \! \! &(M(t+\tau) \! - \! M(t))^H X(1) \nonumber\\
   \label{compound1}
   & \mathop{=} \limits_{law} & M(\tau)^H X(1)
\end{eqnarray}
In other words, the variations of $M(t)$ are
related to the local variance of a Brownian motion. In finance, $X(t)$ represents some asset price (or the logarithm
of an asset price) whose increments are the so-called asset returns. In that case, the measure $M(t)$ is usually referred
to as the ``trading time'' or the ``volatility process'' since its increments $M(t+\tau)-M(t) \geq 0$
simply correspond to the volatility
(i.e. the local variance) between times $t$ and $t+\tau$.
Henceforth, most of our considerations will concern the ``volatility''
$M(t)$. All the results can be extended to the ``price'' process $X(t)$
in a straightforward manner using Eq. \eqref{compound1}.

In a loose mathematical sense, a non-decreasing
stochastic process $M(t)$ is called multifractal
(or ``multifractal measure'') if the moments of its increments (assumed to be stationary)
$\delta_\tau M (t) = M(t+\tau)-M(t)$ verify
the multiscaling properties:
\begin{equation}
\label{mscaling}
   \EE{\delta_\tau M(t)^q} = \EE{M(\tau)^q} \sim C_q \tau^{\zeta(q)} \; ,
\end{equation}
where $\zeta(q)$ is a nonlinear concave function of the moment order $q$.
Notice that the multifractal nature is properly defined by the nonlinearity
of $\zeta(q)$ as opposed to monofractal situations where $\zeta(q)$ is a linear
function. In order to quantify the multifractality, one often defines
the so-called {\em intermittency coefficient} $\lambda^2$ as the curvature
of $\zeta(q)$ around $q=0$:
\begin{equation}
  \lambda^2 = -\zeta''(0) \geq 0\; .
\end{equation}
The last inequality simply comes from the concavity of the $\zeta(q)$ spectrum. Indeed,
the scaling behavior of Eq. \eqref{mscaling}
is generally interpreted in the limit of small time scales $\tau \rightarrow 0$.
Accordingly, if one computes for example the kurtosis behavior,
\begin{equation}
\label{kurt}
  {\cal F}(\tau) = \frac{\EE{M(\tau)^4}}{\EE{M(\tau)^2}^2} \sim \tau^{\zeta(4)-2\zeta(2)}
\end{equation}
one directly sees that, because ${\cal F}(\tau) \geq 1$, one must have $\zeta(4) \leq 2 \zeta(2)$.
As shown in Appendix \ref{app:integralscale}, this kind of argument can be generalized (thanks to H\"older inequality) to prove that $\zeta(q)$
is concave. Thanks to Eq. \eqref{compound1}, one can conclude that the increment probability density functions (pdf) of $X(t)$
(the price returns in empirical finance)
cannot keep a constant shape at different time scales $\tau$ (that would be gaussian in the monofractal
situation). It necessarily becomes more and more leptokurtic as
$\tau \rightarrow 0$. Both multiscaling and increasing flatness at small scales are two well known stylized
facts characterizing the return time series of many different financial markets \cite{gh96,ManCalFish,MuDeBa00}.

Let us remark that the previous argument can also be used to show that the scaling \eqref{mscaling}
cannot hold for arbitrary large scales $\tau$. Indeed, since ${\cal F}(\tau) > 1$, if $\zeta(4)-2\zeta(2) < 0$
then Eq. \eqref{kurt} can be valid only on a bounded range of scales. Therefore there exists
an {\em integral scale} $T$ below which multiscaling holds and beyond which one observes monofractal scaling properties
(see Appendix \ref{app:integralscale}).

\subsection{Continuous cascades}
Explicit constructions of multifractal measures can be naturally obtained within the
framework of random cascades. The picture of a random cascade comes from the physics of turbulence
where kinetic energy injected in the flow at some large scale is transferred to the finest scales by successive
steps of eddy fragmentation \cite{BF}. The large scale where the cascade ``starts''
corresponds precisely to the integral scale introduced previously.
Accordingly, a discrete random cascade can be constructed as follows:
one starts with an interval of length
$T$ where the measure $M(dt)$ is uniformly spread (meaning that the density is constant)
and splits this interval in two equal
parts: On each part, the density is multiplied by (positive) i.i.d. random factors $W$.
Each of the two sub-intervals is again cut in two equal parts and the process is repeated infinitely.
Given the discrete and non-stationary nature of such constructions and the fact that they are only defined
in a fixed bounded interval (of size $T$), more recently, continuous cascade constructions have been
proposed. These models can be viewed as a ``densification'' of the discrete construction
\cite{SchMar01,MuBa02,BacKozMuz06} where the multiplication along
the dyadic tree associated with successive fragmentation steps,
\[
dM = \prod_i W_i = e^{\sum_i \ln(W_i)} \; ,
\]
is replaced by the exponential of an integral (instead of a discrete sum) of a white noise (instead
of $\ln(W)$) over a cone-like domain in the time-scale plane (instead of the tree-node set).
More precisely, one defines
\cite{MuBa02,BaMu03}:
\begin{equation}
    dM_{\ell,T}(t) = M_{\ell,T}([t,t+dt]) = e^{\omega_{\ell,T}(t)} dt
\end{equation}
with
\begin{equation}
  \omega_{\ell,T}(t) = \mu_{\ell,T}+\int_{(u,s) \in C_{\ell,T}(t)} dW(u,s)
\end{equation}
where $\mu_{\ell,T}$ is a constant such that $\EE{e^{\omega_{\ell,T}(t)}} = 1$,
$dW(u,s)$ is a white noise associated with some infinitely divisible law (more precisely
an ``independently scattered random measure'' \cite{BaMu03})
and $C_{\ell,T}(t)$ is the cone like domain \footnote{Let us remark that the construction we consider here is a ``causal version''
of the original construction proposed in Refs \cite{MuBa02,BaMu03} where a symmetrical cone was used. All the results and computations
remain unchanged for both constructions.} :
\begin{equation}
\label{defcone}
  (u,s) \in C_{\ell,T}(t) \Longleftrightarrow \{s \geq \ell, t-\min(s,T) \leq u \leq t \}
\end{equation}
\begin{figure}[h]
\rotatebox{270}{
\includegraphics[height=8cm]{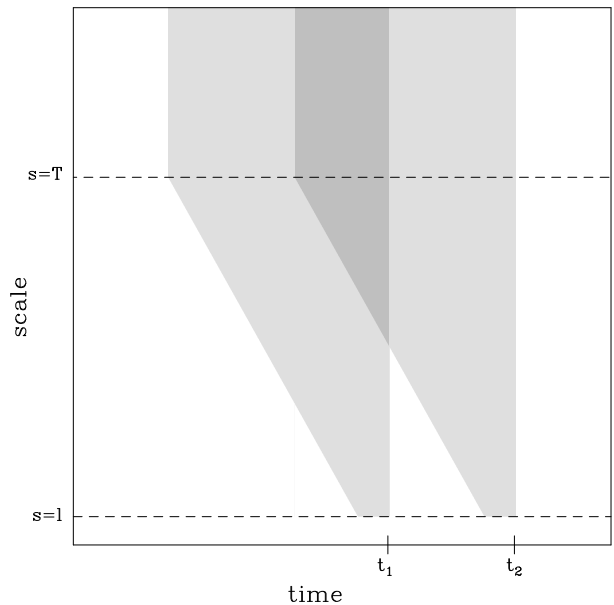}
}
\caption{Construction of a continuous cascade process: $\omega_{\ell,T}(t)$ is the integral
of a white noise over a cone-like domain $C_{\ell,T}(t)$ in the time-scale plane. The covariance of $\omega_{\ell,T}(t_1)$
and $\omega_{\ell,T}(t_2)$ corresponds to the area of the intersection $C_{\ell,T}(t_1) \cap C_{\ell,T}(t_2)$.}
\label{fig1}
\end{figure}
This construction is illustrated in Fig. \ref{fig1}.
The final multifractal measure $dM_T$ is then obtained as the weak limit of $dM_{\ell,T}$ when $\ell \rightarrow 0$, i.e.,
\begin{equation}
\label{defM}
    M_{T}(t) = \lim_{\ell \rightarrow 0} \int_0^t dM_{\ell,T}(t) = \lim_{\ell \rightarrow 0} \int_0^t e^{\omega_{\ell,T}(t)} dt
\end{equation}

For the sake of simplicity, we will consider, in this paper exclusively log-normal
random cascades. All our results can be easily extended to arbitrary log-infinitely divisible laws
within the framework introduced in Refs \cite{MuBa02,BaMu03}.
In the log-normal case, $dW(t,s)$ is a 2D Gaussian (Wiener) white noise of variance $\lambda^2 s^{-2} dt ds$
and it is easy to see (see Fig. \ref{fig1}) that the covariance of $\omega_{\ell,T}(t_1)$ and
$\omega_{\ell,T}(t_2)$ is simply the area of $C_{\ell,T}(t_1) \cap C_{\ell,T}(t_2)$.
Its expression reads:
\begin{equation}
\label{covomega}
   \Cov{\omega_{\ell,T}(t)}{\omega_{\ell,T}(t+\tau)} = \left\{ \begin{array}{ll}
    \lambda^2 \ln(\frac{T}{\tau}) \; \; \;   \mbox{if} \; \ell \leq \tau \leq T \\
    \lambda^2 \left(\ln(\frac{T}{\ell})+ 1-\frac{\tau}{\ell}\right)  \;  \mbox{if} \; \tau \leq \ell \\
    0 \; \; \; \; \;   \mbox{if} \; \tau > T
\end{array} \right.
\end{equation}
In that respect the mean value of $\omega_{\ell,T}$ has to be chosen as:
\begin{equation}
\label{valmu}
   \mu_{\ell,T} = -\frac{\lambda^2}{2}\left(1+\ln(\frac{T}{\ell})\right) \; .
\end{equation}
Notice that in the log-normal case, the intermittency coefficient $\lambda^2$ and the integral scale $T$ are
the only parameters that govern the multifractal statistics.
The previous equation mainly says that the logarithm of a random log-normal multifractal measure
is a Gaussian process which covariance decreases as a logarithmic function, $\log(T/\tau)$.
This features has been shown to directly reflect
the tree structure of discrete random cascades (see Refs \cite{AMS98,ArBaMaMu98}).

\subsection{Stochastic self-similarity}
All the (multi-)scaling properties of $M(t)$ (and subsequently of $X(t)$) can be shown
to result from the logarithmic nature of this covariance.
Indeed, since $\omega_{\ell,T}(t)$ is a Gaussian process,
one can directly infer from Eqs. \eqref{covomega} and \eqref{valmu} that, $\forall \; r < 1$, $\forall \; t \leq T$,
\begin{equation}
\label{selfsim}
\omega_{r\ell,T}(r t) \operatornamewithlimits{=}_{law} \omega_{\ell,T}(t) + \Omega_r
\end{equation}
where $\Omega_r$ is a normal random variable of variance $-\lambda^2 \ln(r)$ and mean $\frac{\lambda^2}{2} \ln(r)$.
From Eq. \eqref{selfsim}, the stochastic self-similarity property results \cite{MuBa02,BaMu03}:
\begin{equation}
\label{stoss}
    M_T(rt) \operatornamewithlimits{=}_{law} r e^{\Omega_r} M(t)
\end{equation}
which directly proves the multiscaling (Eq. \eqref{mscaling}) of the moments of $M(t)$ (and thus of $X(t)$)
with a parabolic $\zeta(q)$ function:
\begin{equation}
  \zeta(q) = q+ \frac{\ln \EE{e^{q \Omega_r}}}{\ln r} = q(1+\frac{\lambda^2}{2}) - \frac{\lambda^2 q^2}{2} \; .
\end{equation}
One can establish another self-similarity property \cite{Al06} when one
also rescales the integral scale. In that case, one has trivially from Eq. \eqref{covomega}, $\forall \; r >0$:
\begin{eqnarray}
\label{selfsimT}
\omega_{r\ell,rT}(r t) & \mathop{=} \limits_{law} & \omega_{\ell,T}(t) \\ \label{stossT}
 M_{rT}(rt) & \mathop{=} \limits_{law} & r M(t)
\end{eqnarray}
which means that a trivial scaling is obtained when the integral time $T$ is rescaled
with the time.

In the field of empirical finance, random cascades have allowed one to understand
that the observed multiscaling properties of return moments and the long-range correlated
nature of the volatility are the two faces of the same coin. The (log-normal) multifractal random
walk model has proven to be a simple, parcimonious model that reproduces most of observed
statistical properties of asset returns \cite{MuDeBa00,BBoPo,durova08,BacKozMuz06}.
As far as statistical estimation issues are concerned, as shown in Ref. \cite{MuPoBa10}, intermittency exponent estimations
based on Eq. \eqref{covomega} are far more reliable than those based on moment multiscaling \eqref{mscaling}
(see also \cite{BacKozMuz11,duv11} for additionnal results on the intermittency exponent estimation using GMM methods).
Empirical evidence for the logarithmic nature of log-volatility correlations
have been provided for different asset price time series over different
markets \cite{AMS98,MuDeBa00,BacKozMuz06,BacKozMuz11,mlmrw}. All these results confirm the multifractal nature of
asset return fluctuations with an intermittency coefficient $\lambda^2 \in [0.01,0.03]$.
However the reported values of the integral scale $T$ vary in a wide range of scales, between few months
and several years. The main question we want to address in this paper concerns that point:
what is the value of the integral scale in financial time series ?

\section{The limit of infinite integral scale: a non-stationary model for log-volatility}
\label{sec_th}
\subsection{Definition of the model}
The broad range of observed values of the integral scale in empirical studies leads us to ask the question
of the interpretation of the integral scale value in financial markets. Unlike turbulence, there is no obvious
large scale that could be singularized and associated with some ``source'' of randomness. Even if the heterogeneity
of agents and the wide range of time horizons used by market participants is a well recognized fact, this can hardly
be invoked to define a single scale that could be as large as several years.

A way to answer the previous remarks could be to consider the model introduced in \cite{SaSo06} where
the authors replaced the log-correlated $\omega_{\ell,T}(t)$ by a long-range (e.g. a fGn) correlated
stationary Gaussian process. However the continuous time limit of such a process is trivial (i.e., it
necessarily involves a small scale cut-off) and its scaling properties are not exact and hard to handle.
Another solution is to define a random cascade process in the limit $T \rightarrow \infty$. However,
the definition of such a limit is not obvious, since,
as emphasized in the previous section and shown in Appendix \ref{app:integralscale},
one cannot define any multiscaling behavior without involving a finite integral scale.
As one can check in Eq. \eqref{covomega}, by letting $T \rightarrow \infty$, one obtains
an infinite value of the variance (and the mean) of $\omega_{\ell,T}$.
In Ref. \cite{durova08}, the authors have considered the possibility of an infinite integral scale
and provided an explicit prediction formula of $\omega_{\ell,T \rightarrow \infty}$ (that we denote as $\omega_{\ell,\infty}$).
However this process is not defined in a classical sense but only in some
quotient space, namely a space of processes defined up to constant time functions.
It has been shown that
\begin{equation}
\label{ws}
    \lim_{T \rightarrow \infty} \int \phi(u) \; \omega_{\ell,T}(t-u) du
\end{equation}
is meaningful for a class of smooth functions $\phi$ provided it is of zero mean.
We already know that the singularity of the covariance function at $\tau = 0$ when $\ell \rightarrow 0$ (Eq. \eqref{covomega})
means that the limit of $\omega_{\ell,T}$ (or $\exp(\omega_{\ell,T})$)
has to be considered as a noise process and
is well defined only when interpreted in the weak (distribution) sense.
When $T \rightarrow +\infty$, Duchon et al. \cite{durova08} show that $\omega_{\ell \rightarrow 0,\infty}$
can be still interpreted in a weak sense but only for test functions satisfying $\int \phi(t) dt = 0$.
This process and notably its exponential $e^{\omega_{\ell,\infty}}$, is however hard to interpret
and of unclear practical interest in quantitative finance.

\begin{figure}[h]
\rotatebox{270}{
\includegraphics[height=8cm]{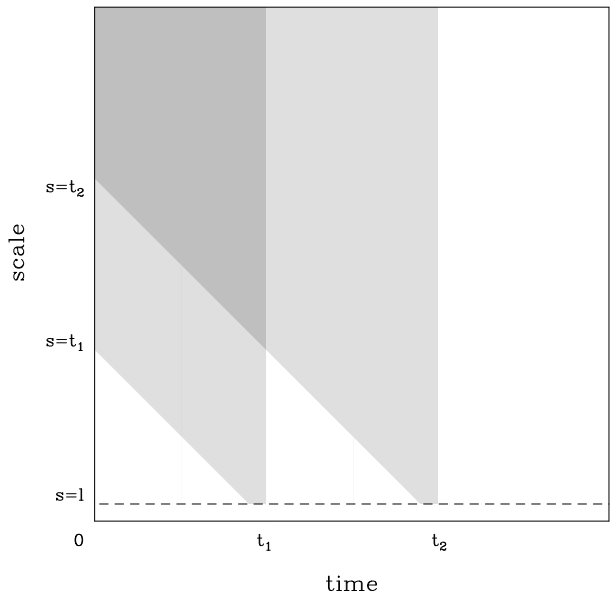}
}
\caption{Construction of the non-stationary $\omega_\ell(t)$ process as the integral
of a white noise over a cone-like domain $C_{\ell}(t)$ in the time-scale plane. The covariance of $\omega_{\ell}(t_1)$
and $\omega_{\ell}(t_2)$ is simply the area of the intersection $C_{\ell}(t_1) \cap C_{\ell}(t_2)$}
\label{fig2}
\end{figure}

In order to handle the low-frequency problem related to $T \rightarrow +\infty$, we propose in this paper
an alternative solution that consists in
considering a {\em non-stationary process} where, at time $t$, the integral
scale is precisely $T=t$.
We define a process $\omega_\ell(t)$ as for standard cascade, from the integration
over a cone-like domain in a time-scale plane, where, at time $t$, the parameter
$T$ in Eq. \eqref{defcone}, is replaced by $t$:
\begin{eqnarray}
(u,s) \in C_{\ell}(t) & \Leftrightarrow & \{s \geq \ell, \max(0,t-s) \leq u \leq t \}  \\ \; \nonumber & \mbox{if} & \; t \geq \ell \\
   C_{\ell}(t) & = & \varnothing \; \mbox{otherwise} \; .
\end{eqnarray}
The process $\omega_\ell(t)$ is then defined by:
\begin{equation}
\label{defomega}
     \omega_\ell(t) = \mu_\ell(t)+\int_{(u,s) \in C_{\ell}(t)} dW(u,s)
\end{equation}
where $\mu_\ell(t)$ is a deterministic mean value defined below and $dW(u,s)$ a Gaussian
white noise of variance $\lambda^2 s^{-2} du ds$.
This construction is illustrated in Fig. \ref{fig2}.
\begin{figure}[h]
\rotatebox{270}{
\includegraphics[height=8cm]{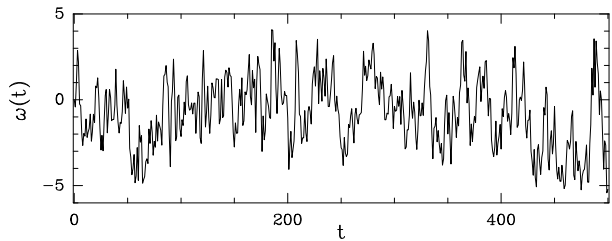}
}
\caption{An example of a path of $\omega_\ell(t)$ where the numerical construction has
been performed by sampling both space and scale in Eq. \eqref{defomega}.}
\label{fig3}
\end{figure}

In Fig. \ref{fig3} we have plotted a sample of $\omega_\ell(t)$ generated at rate $\Delta t = \ell = 1$
over 500 points. As one can see, the non-stationary nature of $\omega_\ell(t)$ is not
obvious (see below).

We can compute the covariance of $\omega_\ell(t)$ that corresponds
to the domain $C_\ell$ intersection areas (see Fig. \ref{fig2}). For $t_1 \leq t_2=t_1+\tau$, its expression reads:
\begin{equation}
\label{covomegans}
\Cov{\omega_{\ell}(t_1)}{\omega_{\ell}(t_2)} = \left\{ \begin{array}{ll}
    \lambda^2 \ln(\frac{t_2}{\tau}) \; \; \;   \mbox{if} \; \tau > l \\
    \lambda^2 \left(\ln(\frac{t_2}{l})+ 1-\frac{\tau}{l}\right)  \;  \mbox{if} \; \tau \leq l \\
    0 \; \; \; \; \;   \mbox{if} \; t < l
\end{array} \right.
\end{equation}
This equation implies notably that
\begin{equation}
\label{varomega}
   \Var{[\omega_\ell(t)]} = \lambda^2\left(1+\ln(\frac{t}{\ell})\right) \; .
\end{equation}
One clearly sees that $\omega_\ell(t)$ is a non-stationary gaussian process but its
covariance has striking similarities with the stationary situation (Eq. \eqref{covomega})
where the integral scale has been replaced by the current time $t$ (or $\max(t_1,t_2)$
in the covariance expression).
\begin{figure}[h]
\rotatebox{270}{
\includegraphics[height=7cm]{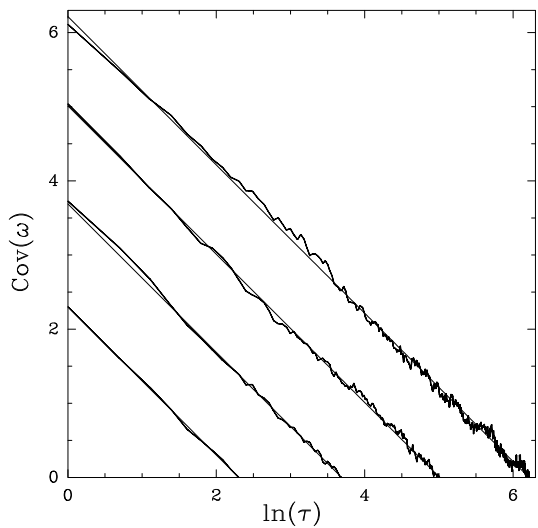}
}
\caption{Covariance function $\Cov{\omega_{\ell}(t_1)}{\omega_{\ell}(t_2)}$
as a function of $\ln(\tau)$, with  $\tau = |t_2-t_1|$ and
$t_2=10, 40, 150, 500$ (from bottom to top curves).
The bold lines correspond to numerical estimates using 500 samples of $\omega_\ell(t)$ while
the thin lines correspond to the analytical expressions (Eq. \eqref{covomegans}). We have chosen
$l=1$ and $\lambda^2 = 1$.}
\label{fig4}
\end{figure}

\begin{figure}[h]
\rotatebox{0}{
\includegraphics[height=7cm]{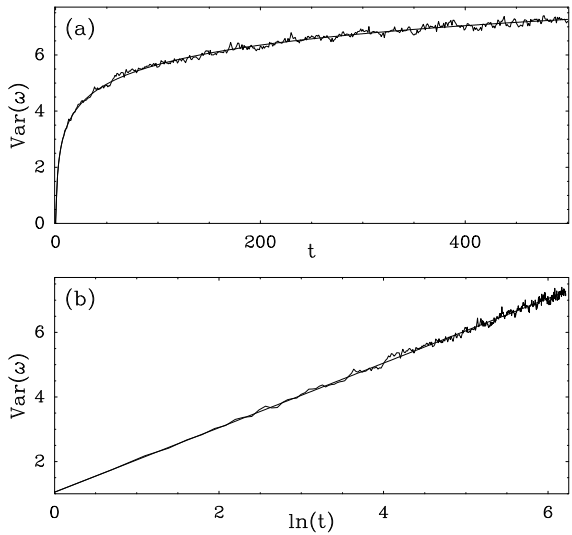}
}
\caption{Variance $\Var{\omega_\ell(t)}$ as a function of $t$ (a) and $\ln(t)$ (b).
We have superimposed to the expected analytical expressions \eqref{varomega},
the estimated variance using 500 Monte-Carlo samples of $\omega_\ell(t)$ with
$l = 1$ and $\lambda^2 = 1$.}
\label{fig5}
\end{figure}

The non-stationary behavior of the covariance is illustrated in Fig. \ref{fig4} where
we have plotted the estimated as well as analytical $\Cov{\omega_{\ell}(t_1)}{\omega_{\ell}(t_2)}$
as a function of $ \ln(\tau) = \ln(t_2-t_1)$ for different
values of $t_2$.
Remark that this kind of non-stationarity is reminiscent of an {\em aging} behavior
as observed in off-equilibrium relaxing systems \cite{aging1,aging2} where
the ``age'' of the process $t_2$ controls the characteristic correlation length.
The logarithmic behavior of the variance is illustrated in Fig. \ref{fig5}.

Let us show that one can choose a function $\mu_\ell(t)$ in Eq. \eqref{defomega} such that one can define the limit
$\ell \rightarrow 0$ of $e^{\omega_\ell}$
in the weak sense, i.e.,:
\begin{equation}
\label{defM}
   M(t) = \lim_{\ell \rightarrow 0} \int_0^t e^{\omega_\ell(u)} du
\end{equation}
In fact, as for continuous stationary cascades,
one can use a general argument on positive martingales (as e.g., in Ref. \cite{BaMu03})
if, for all time interval $I$, $\int_I e^{\omega_\ell(t)} dt$ is a martingale as a function of $\ell$. This is precisely the case provided,
$\forall \; t$,
\[
 \EE{e^{\omega_\ell(t)}} = 1
\]
a condition equivalent, in the log-normal case, to
\begin{equation}
\label{relmuv}
   \mu_\ell(t) = -\frac{1}{2} \Var{[\omega_\ell(t)]} = -\frac{\lambda^2}{2}\left(1+\ln(\frac{t}{\ell})\right) \; .
\end{equation}
In Appendix \ref{app:msconv}, we provide an alternative direct proof of mean square convergence.

Notice that this equation also guarantees that
\begin{equation}
 \EE{M(t)} = \Var{[X(t)]} = \sigma^2 t
\end{equation}
(recall that $X(t) = B[M(t)]$ with $B(t)$ a standard Brownian motion).

\subsection{Scaling and self-similarity properties}
Let us remark that increments of $\omega_\ell(t)$, $\delta_h \omega_\ell(t) = \omega_\ell(t+h)-\omega_\ell(t)$
($h > \ell$) have a time dependent variance so they are not stationary. However,
for $\tau > h$, their covariance depends only on the lag $\tau$. After a little algebra,
their expression reads:
\begin{equation}
\label{inc_cov}
\Cov{\delta_h\omega_{l}(t)}{\delta_h\omega_{l}(t+\tau)} = \lambda^2 \ln\left(1-\frac{h^2}{\tau^2}\right) \; .
\end{equation}
This covariance function corresponds to a power-spectrum such that $P_{\delta_h \omega}(f) \sim |f|$ when $f \ll h^{-1}$.
Since $P_{\omega_\ell}(f) \sim f^{-2} P_{\delta_h \omega_\ell}(f)$,
it results that $\lim_{\ell \rightarrow 0} \omega_\ell(t)$ can be associated with
a $1/f$ power-spectrum. Let us mention that, in Ref. \cite{aging2}, the author has already raised the possibility
of an ``aging'' non-stationary model in order to handle the low-frequency problem of $1/f$ noise.
In that respect, $\omega_{\ell \rightarrow 0}(t)$ can be interpreted as the limit $H \rightarrow 0$ of a fractional
Brownian motion (fBm) $B_H(t)$ of Hurst parameter $H$ \cite{TaqSam94}.

This interpretation of $\omega_{\ell}(t)$ can also be
suggested from its self-similarity properties.
Indeed, from the covariance expression \eqref{covomegans},
one can establish the following invariance relationship for $\omega_\ell(t)$:
\begin{equation}
\label{selfsim2}
\omega_{r \ell}(r t) \operatornamewithlimits{=}_{law} \omega_{\ell}(t) \; .
\end{equation}
This equality extends to $H=0$ the standard fBm self-similarity
$B_H(rt) =_{law} r^{H} B_H(t)$ \footnote{Remark that, since fBm's are defined from both their self-similarity
properties and the stationarity of their increments \cite{TaqSam94}, in full rigor $\omega_{\ell \rightarrow 0}(t)$
cannot be identified to $B_H(t)$ with $H=0$.}.
It is noteworthy that $\omega_{\ell \rightarrow 0}(t)$ has the drawbacks of both fractional Gaussian noises
and fractional Brownian motions
since it exists only in the sense of distributions and it is a non-stationary process.
From the definition \eqref{defM} and thanks to previous equality,
one deduces the simple self-similarity property of the volatility process $M(t)$:
\begin{equation}
\label{stoss2}
    M(rt) \operatornamewithlimits{=}_{law} r M(t) \; .
\end{equation}
Let us remark that relation \eqref{selfsim2} is different from Eq. \eqref{selfsim} but
can be understood as reminiscent of Eq. \eqref{selfsimT} where one allows
the integral scale to become infinite (i.e.,
$T \rightarrow \infty$).

When one compares the self-similarity of $M$ and $M_T$ (Eqs. \eqref{stoss2} and \eqref{stoss}), one sees that
in the former case there is no stochastic factor $e^{\Omega_r}$ and the scaling of the moments of $M$ (and therefore of the return process $X(t)$) becomes
trivial:
\begin{equation}
   \EE{M(\tau)^q} = C_q \tau^q \Rightarrow \EE{|X(\tau)|^q} = K_q \tau^{\frac{q}{2}} \; .
\end{equation}
In the sense of Eq.~\eqref{mscaling}, it thus appears that $M(t)$ (or $X(t)$) is
not a multifractal process. However, one must carefully interpret the previous equation
since $M(t)$ (and then $X(t)$) has no stationary increments.
It results that there is no reason that the moments $\EE{M(\tau)^q}$ and
$\EE{[M(t+\tau)-M(t)]^q}$ behave in the same way. Let us make the explicit computation
for $q=2$.
In that case,
\begin{eqnarray*}
\EE{M(\tau)^2} & = & \lim_{\ell \rightarrow 0} \int_0^\tau \int_0^\tau \EE{e^{\omega_\ell(u)+\omega_\ell(v)}} du dv \\
               & = &  \lim_{\ell \rightarrow 0} \int_0^\tau \int_0^\tau e^{\Cov{\omega_\ell(u)}{\omega_\ell(v)}} du dv \\
               & = & \int_0^\tau \int_0^\tau \left(\frac{\max(u,v)}{|u-v|}\right)^{\lambda^2} du dv \\
               & = & \tau^2 \int_0^1 \int_0^1 \left(\frac{\max(u,v)}{|u-v|}\right)^{\lambda^2} du dv \\
               & = & C_2 \tau^2
\end{eqnarray*}
whereas,
\begin{eqnarray*}
& \EE{[M(t+\tau)-M(t)]^2} =  \\
& \lim_{\ell \rightarrow 0} \int_t^{t+\tau} \int_t^{t+\tau} \EE{e^{\omega_\ell(u)+\omega_\ell(v)}} du dv \\
& = \int_t^{t+\tau} \int_t^{t+\tau} \left(\frac{\max(u,v)}{|u-v|}\right)^{\lambda^2} du dv \\
& = t^2 \int_1^{1+\frac{\tau}{t}} \int_1^{1+\frac{\tau}{t}} \left(\frac{\max(u,v)}{|u-v|}\right)^{\lambda^2} du dv \\
& = t^2 \int_0^{\frac{\tau}{t}} \int_0^{\frac{\tau}{t}} \left(\frac{1+\max(u,v)}{|u-v|}\right)^{\lambda^2} du dv \; .
\end{eqnarray*}
If one supposes that $\frac{\tau}{t} \ll 1$, then in the last integral
the term $\max(u,v) \ll 1$ can be neglected and, using the change of variables $u'= u t/\tau$ and $v' = v t/\tau$,
one gets:
\[
\EE{[M(t+\tau)-M(t)]^2} \simeq C_2(t) \tau^{2-\lambda^2}
\]
where the constant $C_2(t) \sim t^{-\lambda^2}$. The previous equation shows that
in the limit $\tau \ll t$, the mean square of the increments of $M(t)$ behaves
as the increment of the multifractal measure $M_T(t)$ (with the scaling exponent $\zeta(2) = 2-\lambda^2$)
where $t$ plays precisely
the role of the integral scale $T$.
This behavior can be directly established from the expression of the covariance, Eq. \eqref{covomegans}: Indeed, let us
consider two times $t_1,t_2$ in some interval $[t_0,t_0+\Delta t]$. If $\Delta t \ll t_0$, then
to the first order in $t_0/\Delta t$,
we have $\Cov{\omega_\ell(t_1)}{\omega_\ell(t_2)} = \lambda^2 \ln(t_0/|t_1-t_2|)$, i.e. the same
covariance as the process $\omega_{\ell,T}$ used to build an exact multifractal random measure
with $T = t_0$. This means that the non-stationary process $M(t)$ defined in Eq. \eqref{defM}, cannot be distinguished from a (stationary) multifractal random measure $M_{t_0}(t)$ of integral scale $T=t_0$ over any interval $[t_0,t_0+\Delta t]$
(to the first order in $t_0/\Delta t$).


\section{Application to financial data}
\label{sec_apps}
As recalled in the introduction, various authors have suggested that most
of stylized facts characterizing the volatility associated with asset prices in financial markets
can be accounted by multifractal measures. Let us illustrate how the model $M(t)$ introduced in this paper,
allows one to explain the large discrepancies of the reported integral scale values as a consequence of the non-stationary nature
of log-volatility. Since the model is non-stationary and since in practice there is no possibility to have
an ensemble of many independent samples, one has first to discuss which kind of estimation
one can perform on a single realization of the volatility.

\subsection{Pathwise properties and estimation issues}
Let us suppose that one studies a multifractal random measure $M_T(t)$ (i.e. a classical random cascade with finite
integral scale $T$) over an interval
$[t_0,t_0+\Delta t]$ (or, since $M_T$ has stationary increments, over $[0,\Delta t]$) with $\Delta t < T$.
Then from the self-similarity relations \eqref{selfsim} and \eqref{selfsimT}, one as, for all $r < \Delta t/T < 1$,
\begin{equation}
M_{T}(t) \mathop{=} \limits_{law} r^{-1} M_{rT}(rt) \mathop{=} \limits_{law} r^{-1} e^{\omega_r} M_{rT} (t) \; .
\end{equation}
Since the random variable $\omega_r$ is fixed on a single realization, this equality
clearly means that one cannot distinguish over any interval $[t_0,t_0+\Delta t]$
two multifractal measures
$M_{T_1}(t)$ and $M_{T_2}(t)$ with $T_1 \neq T_2$ and $T_1,T_2 \geq \Delta t$. Estimating the integral scale on a single realization
of $M_T(t)$ over an interval of length $\Delta t < T$ is thus impossible. The question is to which
value an empirical estimation leads to ?

Empirically, as advocated e.g., in Ref. \cite{BacKozMuz11},
the correlation properties of $\omega_{\ell,T}(t)$ can be estimated
using a proxy (called the ``magnitude process'') of $\omega_{h,T}(t)$ estimated
from the logarithm of the increments of $M_T(t)$: $\omega_{h,T} \simeq \ln \delta_h M_T(t)$.
If $\omega_{h,T}$ is sampled at rate $h$ over a time period of length $\Delta t$, the estimator of its covariance $\widehat{C_{\Delta t}}(\tau)$ at lag $\tau = n h$, reads:
\begin{equation}
\label{covestdef}
  \widehat{C_{\Delta t}}(\tau) = (N-n)^{-1} \! \! \! \sum_{i=0}^{N-1-n}\!  \! \! (\omega_{h,T}[ih]-{\hat \mu})(\omega_{h,T}[(i+n)h]-{\hat \mu})
\end{equation}
where $N=\frac{\Delta t}{h}$ is the sample size and ${\hat \mu}$ is the empirical mean:
${\hat \mu} = N^{-1} \sum_{k=0}^{N-1} \omega_{h,T}(kh)$.
In Appendix \ref{app:covest} (see also Ref. \cite{BacKozMuz11} for a more technical approach)
it is shown that:
\begin{equation}
\label{covest}
 \EE{\widehat{C_{\Delta t}}(\tau)} \simeq \lambda^2 \left( \ln \left( \frac{e^{-3/2} \Delta t}{\tau} \right) -\frac{\tau}{\Delta t} \right)+O\left(\frac{\tau^2}{\Delta t^2} \right)\; .
\end{equation}
This equation means that, over a sample of size $\Delta t$, the estimated auto-covariance
of the magnitude associated with a multifractal process of integral scale
$T > \Delta t$ is the auto-covariance of a multifractal process of integral
scale $e^{-3/2} \Delta t$.

\begin{figure}[h]
\rotatebox{270}{
\includegraphics[height=7cm]{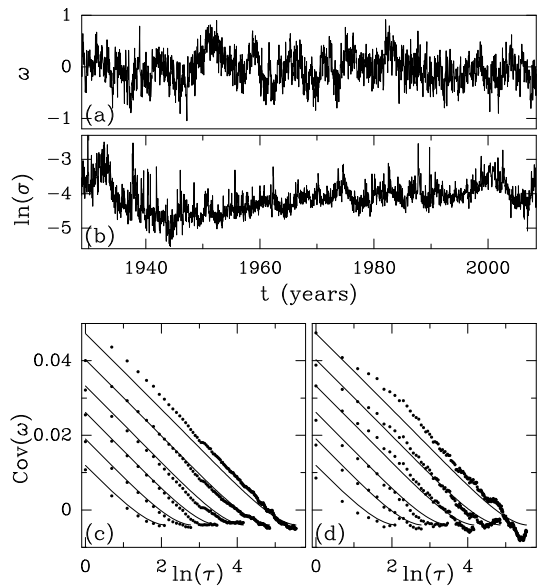}
}
\caption{(a) Sample path of $\omega_h(t)$ of length $2.10^{4}$
where the numerical construction has
been performed by sampling in both space and scale the cone-like sets $C_\ell(t)$
for small $\ell$ in order to define $M(t)$.
(b) Magnitude estimated as $\ln(\sigma(t))$ where $\sigma(t)$ is the daily range
(difference between highest and lowest daily return values) associated with the Dow-Jones
time series from 1929 to 2011.
In (c) and (d) are reported the magnitude auto-covariance estimation in semi-logarithmic scale
for different sample sizes $\Delta t = 16,32,64,128,256$ and $512$
(from bottom to top). (c) Estimation from the model sample path in (a).
(d) Estimation from the Dow-Jones daily data in (b).
The solid lines represent the theoretical predictions from Eq. \eqref{covest} with $\lambda^2 = 0.01$.}
\label{fig6}
\end{figure}

If we now go back to the non-stationary process $M(t)$, since we have shown that,
over every interval $[t_0,t_0+\Delta t]$,
$M(t)=_{law} M_{t_0}(t)$, we can conclude that, as soon as $t_0 > \Delta t$,
the estimated auto-covariance of $\omega_h(t)= \ln [M(t+h)-M(t)]$ will be provided by
Eq. \eqref{covest}. In other words, for observations far from the time origin, the
estimated integral scale is always (up to a constant factor) the overall sample size.
This is illustrated in Fig. \ref{fig6}(c) where we have
reported the estimation of the magnitude auto-covariance for various sample lengths $\Delta t$.
More precisely, we have generated a single large sample of the process $M(t)$ from which
the magnitude time series $\omega_{h}(t)$ has been computed. This series
(of overall size $L = 2. 10^4$) is displayed in Fig. \ref{fig6}(a).
For each subinterval size $\Delta t = 16,32,\ldots,512$, the sample
is divided in $L/\Delta t$ sub-samples
of length $\Delta t$. The reported estimator $\widehat{\C_{\Delta t}}(\tau)$
is the average of the obtained empirical covariances over all of the $L/\Delta t$ intervals.
One can check in Fig. \ref{fig6}(c) that the theoretical predictions \eqref{covest} (solid lines) are, for all $\Delta t$, in good agreement with the observations
($\bullet$) and one clearly observes an apparent integral scale that grows with $\Delta t$ (as $e^{-3/2} \Delta t$).

\subsection{Application to daily stock data}
Let us now apply the previous analysis to real data.
We report below the empirical results we obtained on three stock indices (namely the Dow-Jones,
the CAC40 and the FTSE100 indices)
over sufficient long time periods. In each case, $h=1$ day and $\omega_h(k)$ at day $k$ is estimated
as $\omega_h(k) = \ln (\sigma(k))$, where $\sigma(k)$ is the relative range
computed from highest and lowest stock values observed during the day $k$.
The considered time periods are 1929-2011 for the Dow-Jones series (around 21.000 trading days), 1990 to 2011 for the CAC40 (around 5500 trading days),
1984 to 2011 (around 7000 trading days) for the FTSE100.

\begin{figure}[h]
\rotatebox{270}{
\includegraphics[height=8cm]{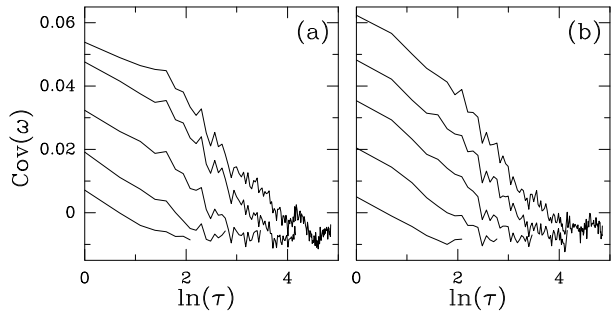}
}
\caption{Magnitude covariance estimation $\widehat{C_{\Delta t}}(\tau)$ as a function of $\ln(\tau)$
for respectively (a) CAC40 (5500 data points) and (b) FTSE 100 (7000 data points)
stock index series. Each curve corresponds to a different sample size $\Delta t$ used for the estimation
($\Delta t = 16, 32, 64, 128, 256$) and for each $\Delta t$, $\widehat{C_{\Delta t}}(\tau)$ has been obtained as the mean
value over all available periods of size $\Delta t$. One clearly sees that the behavior is the same
than for the Dow-Jones index in Fig. \ref{fig6}(d): the integral scale (intercept)
is growing as a function of $\Delta t$. The noise amplitude is greater because the overall sample sizes are smaller than
for the Dow-Jones series.}
\label{fig7}
\end{figure}

\begin{figure}[h]
\rotatebox{270}{
\includegraphics[height=6cm]{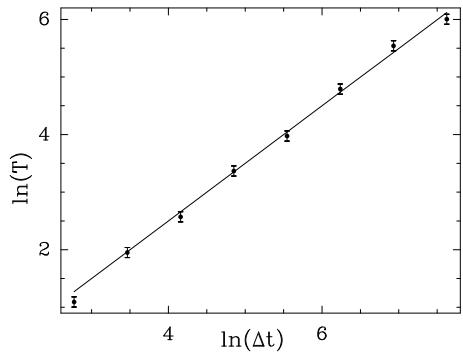}
}
\caption{Estimated integral scale $T$ as a function of the sample size $\Delta t$ (in log-log coordinates)
for the Dow-Jones daily time series from 1929 to 2011. The solid line represents the value $e^{-3/2} \Delta t$
one expects theoretically. The reported error bars correspond to standard deviation of the empirical
mean values estimated from the observed dispersion over all sub-intervals.}
\label{fig8}
\end{figure}

\begin{figure}[h]
\includegraphics[height=5cm]{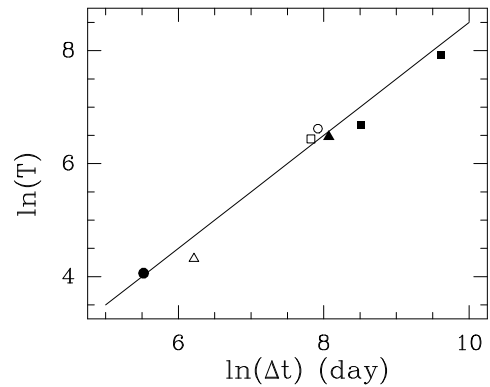}

\caption{Estimated integral scale $T$ as a function of the sample length $\Delta t$ (in log-log coordinates)
gathered from several recent studies in the literature: ($\bullet$) from \cite{turbfin}, ($\circ$) from \cite{BacKozMuz06}, ($\vartriangle$) from
\cite{AMS98}, ($\blacktriangle$) from \cite{MoDiAs12}, ($\square$) from \cite{MuDeBa00} and ($\blacksquare$) from \cite{mlmrw}.
The solid line represents the model fit $\ln(T) = \ln(\Delta t)-3/2$.}
\label{fig9}
\end{figure}

In Fig. \ref{fig6}(b) is plotted the time series corresponding to the daily log-volatility $\omega_h(k)$
of the Dow-Jones index. Very much like the
model (Fig. \ref{fig6}(a)), one can observe excursions away from the mean value lasting for several years.
For each of the 3 volatility series, we reproduced the same covariance estimation experiment we conducted for
the model (Fig. \ref{fig6}(c)).
In Fig. \ref{fig6}(d) are reported the results obtained for the Dow-Jones index while in Fig \ref{fig7} are
reported the results obtained for the CAC40 and the FTSE100 time series.
Tough these latter series have a smaller size and lead to more noisy results, it clearly appears in all
cases that the empirical auto-covariance functions
are fairly well fitted by a multifractal logarithmic shape $\lambda^2 \ln(T/\tau)$ (i.e. they are linear as functions of $\ln(\tau)$)
with a constant intermittency coefficient $\lambda^2 \simeq 0.01$.
However the apparent integral scale $T$ (the intercept of each curve) appears to strongly
depend on $\Delta t$. In Fig. \ref{fig6}(d), we see that the model predictions (solid lines) as described by Eq. \eqref{covest},
fit fairly well the data.
This is confirmed in Fig. \ref{fig8}, where we have plotted (in log-log scale) the estimated integral
scale as a function of the sample size $\Delta t$. One can see that
the analytical prediction $T(\Delta t) = e^{-3/2} \Delta t $ (solid line)
is in very good agreement with the empirical data.
These results allow us to understand the origin of the wide range of integral scale values (from few months to several years) reported
in the literature so far. This is
illustrated in Fig. \ref{fig9} where we have reported the estimated values of the integral scale
$T$ gathered from the recent literature \cite{turbfin,BacKozMuz06,AMS98,MoDiAs12,MuDeBa00,mlmrw}.
Even if these studies concern various data sets
at different time resolutions (intradays, daily,..), different time periods and correspond to different asset classes (FX rates, stocks,..), we
see that the reported values of $T$ are spread closely around the theoretical curve (solid line in Fig. \ref{fig9}).

\section{Conclusion and prospects}
\label{sec_cp}
To conclude we have introduced a new model of stochastic measure as the exponential of a non-stationary
gaussian $1/f$ noise. We have shown that, over any finite time interval, provided
the considered time $t$ is large enough, this model can be hardly distinguished
from a multifractal random cascade with an integral scale that is equal to the sample length.
Our approach can be very appealing to model all phenomena where multiscaling properties are observed
without the existence of any natural large ``correlation'' (or ``injection'') scale in space or time.
For example, in finance, the agreement of the model predictions
with the observed behavior of log-volatility correlation in
various stock indices is striking. These findings suggest a peculiar
(aging) non-stationary nature of volatility fluctuations.
The question of the meaning of the time origin, the possibility to estimate this time
from empirical data will have to be considered in future works.
On a more general ground, the explanation of such non-stationarity is an important question
that will have to be addressed from the market dynamical properties
at microstructure level but also within the framework of agent based approaches including
behavioral finance or theory of self-referencing dynamics of market prices.
On a mathematical ground, it will be interesting to study this model and its possible
variants in relationship with fractional Brownian motion, since it offers the possibility
to give a meaning to the limit $H \rightarrow 0$. Finally our approach can shed a new light in the
field of $1/f$ noise modeling.

\appendix
\section{Proof of the concavity of $\zeta(q)$ and the existence of an integral scale}
\label{app:integralscale}
Let us prove that if Eq. \eqref{mscaling} holds in the limit of small time scales $\tau$, then
i) $\zeta(q)$ is a concave function of $q$ and ii) it necessarily involves a bounded scale $T$ below which it can no longer be valid.
We start by assuming that the following scaling holds in some range of scales:
\[
  \EE{|\delta_\tau X(t)|^q} \sim C_q \tau^{\zeta(q)} \;
\]
Let $F(q,\tau) = \ln \left( \EE{|\delta_\tau X(t)|^q} \right)$. Then by H\"older inequality, $F(q,\tau)$ is, for each $\tau$, a convex
function of $q$. If one assumes it is regular enough so that its second derivative exists, one thus has, $\forall \; \tau > 0$:
\begin{equation}
         F''(q,\tau) \geq 0 \; .
\end{equation}
If the previous scaling law holds, this can be written as:
\begin{equation}
           \frac{d^2 \ln C_q}{dq^2} + \zeta''(q) \ln(\tau) \geq 0
\end{equation}
One sees that if the scaling holds in the limit $\tau \rightarrow 0$, this inequality can be true only
if $\zeta''(q) \leq 0$, i.e.,
$\zeta(q)$ must be a concave function of $q$. If the scaling is also valid at scale $\tau = 1$
(up to a redefinition of $\tau$ we can always assume it is the case),
$C_q$ is the order $q$ moment of the random variable $\delta_1 X(t)$ and
$c_q = \frac{d^2 \ln C_q}{dq^2} \geq 0$. This means that:
\begin{equation}
         \ln(\tau) \leq  \frac{c_q}{-\zeta''(q)}  \; .
\end{equation}
In other words, if $\zeta(q)$ is strictly concave (multifractal case), the scaling
can only hold in a limited range of scales and there exists an integral scale
\[
    T = \inf_q  \left(e^{\frac{-c_q}{\zeta''(q)}}\right)
\]
above which it is not valid.

\section{Proof of the mean-square convergence of $M_\ell(t)$}
\label{app:msconv}
Let us provide a direct proof of the mean square weak convergence of $M_\ell(t)$ (or $M_{\ell}(I) = \int_I e^{\omega_{\ell}(u)} du$ for a any given time interval $I$) as
defined in \eqref{defM} when $\ell \rightarrow 0$.
For that purpose let us show that
\begin{equation}
\label{conv}
  \lim_{\ell,\ell' \rightarrow 0} \EE{(M_\ell(t)-M_{\ell'}(t))^2} = 0 \; .
\end{equation}
Without loss of generality, we assume in the sequel that $\ell' \geq \ell$.
Since,
\begin{eqnarray*}
& \EE{(M_\ell(t)-M_{\ell'}(t))^2} = \EE{\left(\int_0^t \left(e^{\omega_\ell(u)}-e^{\omega_{\ell'}(u)} \right) du \right)^2} \\
= \! & \! \int_0^t \!\int_0^t du \; dv \; \EE{e^{\omega_\ell(u)+\omega_\ell(v)}+e^{\omega_{\ell'}(u)+\omega_{\ell'}(v)} \! \! - \! 2 e^{\omega_\ell(u)+\omega_{\ell'}(v)}}
\end{eqnarray*}
and since $(\omega_\ell,\omega_{\ell'})$ is a vector of correlated Gaussian processes,
thanks to Eq. \eqref{relmuv}, $\EE{(M_\ell(t)-M_{\ell'}(t))^2}$ reduces to:
\begin{equation}
\label{iii}
 \int_0^t \int_0^t du \; dv \left[e^{C_{\ell,\ell}(u,v)}-e^{C_{\ell',\ell'}(u,v)} \right]
\end{equation}
where we denoted $C_{\ell,\ell'}(u,v) = \Cov{\omega_{\ell}(u)}{\omega_{\ell'}(v)}$
and used the obvious property $C_{\ell,\ell'}(u,v) = C_{\ell',\ell'}(u,v)$ if $\ell' \geq \ell$.
Let us split the integral in 3 domains:
\[
\int_0^t \int_0^t = \int\int_{|u-v| \leq \ell} + \int\int_{\ell \leq |u-v| \leq \ell'} + \int\int_{|u-v| \geq \ell'} \; .
\]
It is clear that in the last interval, $C_{\ell,\ell}(u,v) =  C_{\ell',\ell'}(u,v) = \lambda^2 \ln(\frac{\max(u,v)}{|u-v|})$. The corresponding integral in Eq. \eqref{iii} is thus zero.
In interval $\ell \leq |u-v| \leq \ell'$, thanks to expression \eqref{covomegans}, one has
\[
\int \! \int_{\ell \leq |u-v| \leq \ell'} \left(e^{C_{\ell,\ell}(u,v)}- e^{C_{\ell',\ell'}(u,v)}\right) \; du dv = O(\ell'^{1-\lambda^2})
\]
while in the last interval,
\[
\int\int_{|u-v| \leq \ell} e^{C_{\ell,\ell}(u,v)}- e^{C_{\ell',\ell'}(u,v)} = O(\ell^{1-\lambda^2})
\]
proving the mean square convergence \eqref{conv}.

\section{Magnitude covariance estimation}
\label{app:covest}
Let us establish Eq. \eqref{covest}.
Let us denote ${\widehat C(n)} = {\widehat C_{\Delta t}(\tau=nh)}$ and $C(n) = \lambda^2 \ln(T/nh)$ the theoretical covariance
as given by Eq. \eqref{covomega} at lag $\tau = nh$.
By taking the expectation of expression \eqref{covestdef} after expanding the expression of ${\hat \mu}$, one
finds:
\begin{equation}
\label{eqC}
  \EE{{\widehat C(n)}} = C(n)+K(0)-2 K(n)
\end{equation}
where
\[
  K(n) = \frac{1}{N(N-n)}\sum_{i=0}^{N-n-1} \sum_{j=0}^{N-1} C(|i-j|) \; .
\]
If $h$ is small enough, one can replace the double sum by its integral approximation:
\[
   K(n=\tau/h) = \frac{\lambda^2}{\Delta t^2(1-\frac{\tau}{\Delta t})} \int_0^{\Delta t} \int_0^{\Delta t-\tau} \! \! du \; dv \ln\left( \frac{T}{|u-v|} \right) \; .
\]
Evaluating this integral leads, to the first order in $\tau/\Delta t$, to the expression:
\[
   K(n=\tau/h) = \lambda^2 \left( \ln \left(\frac{Te^{3/2}}{\Delta t}\right)-\frac{\tau}{2 \Delta t} \right) \; .
\]
Inserting this expression in Eq. \eqref{eqC}, one gets Eq. \eqref{covest}.


\end{document}